%% file: z__main.tex
\documentclass{article}
\usepackage[utf8]{inputenc}

\usepackage[]{authblk}

\usepackage{graphicx}

\usepackage{url}

\title{Advanced Data Protection Control (ADPC):\\ An Interdisciplinary Overview}
\author{Soheil Human}
\affil{Sustainable Computing Lab\\ Institute for Information Systems and New Media\\ Vienna University of Economics and Business (WU Wien)\\ Vienna, Austria\\soheil.human@wu.ac.at}

\date{}

\begin{document}

\maketitle
\input{00_logo.tex}

\input{00_abstract.tex}

\input{01_introduction.tex}

\input{02_problems-and-necessities.tex}

\input{03_adpc.tex}

\input{06_conclusion.tex}

\input{07_acknowledgement.tex}

\bibliographystyle{unsrt}
\bibliography{references}

\end{document}

%% file: 00_logo.tex
\begin{figure}[h!]
\centering
  \includegraphics[width=0.5\textwidth]{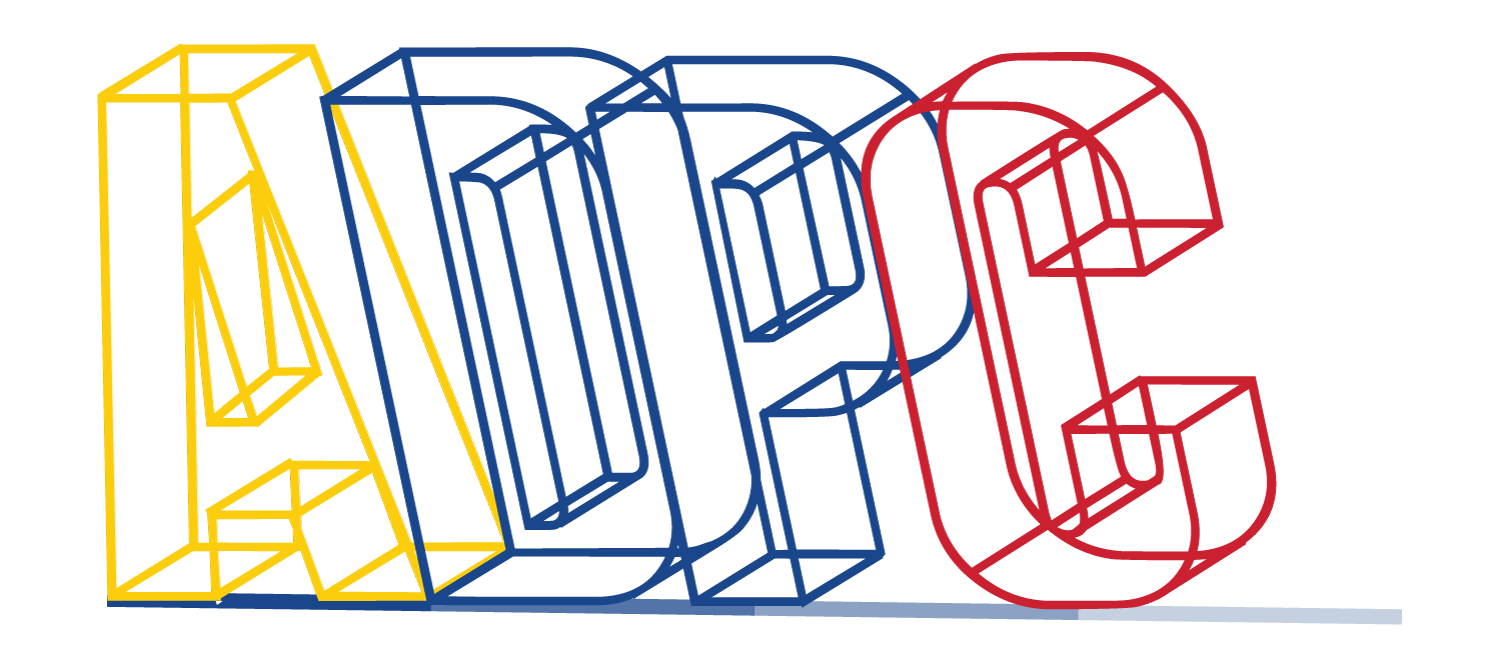}
  \caption{The Logo of the Advanced Data Protection Control (ADPC)}
  \label{fig:ADPC-logo}
\end{figure}

%% file: 00_abstract.tex
\begin{abstract}
The \textit{Advanced Data Protection Control} (ADPC) is a technical \linebreak specification --- and a set of sociotechnical mechanisms surrounding it --- that can change the current practice of Internet-based personal data protection and consenting by providing novel and standardized means for the communication of privacy and consenting data, meta-data, information, requests, preferences, and decisions.
The ADPC supports humans in practicing their rights to privacy and agency by giving them more human-centric control over the processing of their personal data and consent. It helps the data controllers to improve their users' experiences and provides them with easy-to-adopt means to comply with the relevant legal and ethical requirements and expectations.

\end{abstract}

%% file: 01_introduction.tex
\section{Introduction}
\label{sec:introduction}

The protection of personal data is a human right \cite{fuster2014emergence, rodota2009data}. 
Therefore, personal data ``must be processed fairly for specified purposes and on the basis of the consent of the person concerned, or some other legitimate basis laid down by law'' (Art. 8, Charter of Fundamental Rights of the European Union). Accordingly, empowering \cite{2020__AltEtAlEndUserEmpowerment, 2020__HumanEtAlEndUserEmpowerment} data subjects (e.g., users, customers), data controllers \cite{burmeisterPrivacydrivenEnterpriseArchitecture2019} (e.g., companies, organizations, service providers) and other involved actors to co-construct Human-centric, Accountable, Lawful, and Ethical (HALE) \cite{humanHALEWHALEFramework2022} data protection and consenting practices through sociotechnical systems should be considered an essential means for keeping (or making) our societies \linebreak sustainable~\cite{humanHumancentricPersonalData2022}.

\paragraph{Problem:} 
In recent years, a vast amount of research has shown that we have failed to protect end-users' right to privacy and agency (see, e.g., \cite{matte2020cookie, zuboff2015big, isaak2018user}). The consequences of this failure go way beyond the invasion of \textit{individuals' privacy} and can negatively influence different aspects of our societies \cite{2019__HumanEtAl[How]CanPluralist}. Therefore, tackling \textit{the problem of digital privacy and agency} in a human-centric manner is a fundamental challenge of our time to protect humans' rights and keep/make our societies \textit{digitally sustainable} \cite{2021__HumanEtAlHumanCentricitySustainable}.

While data controllers ``control'' most of the fundamental aspects of online privacy and consenting (see Section~\ref{sec:necessity}), data subjects are mostly left alone to deal with the difficulties and complexities of managing (and protecting) their online privacy and agency --- a \textit{mission impossible}.

The Advanced Data Protection Control (ADPC\footnote{The Figure \ref{fig:ADPC-logo} shows the logo of the ADPC.}) aims to tackle this important problem by contributing to a fundamental shift in the practice of online privacy and agency through enabling new communication mechanisms and, consequently, new \textit{assistive solutions} --- and ultimately by changing the control and power structures. 

\paragraph{Objective:} This paper presents the Advanced Data Protection Control (ADPC) \cite{humanAdvancedDataProtection2021}, 
a mechanism for the communication of data protection and consenting data, meta-data, information, requests, preferences, and decisions that can contribute toward a fundamental shift in the ways that personal data processing is practiced on the Internet: i.e., from a \textit{data-controller-centric} practice to a \textit{data-subject-centric} (i.e., a \textit{human-centric}) practice. The aim here is to provide an interdisciplinary overview that is understandable for experts across different disciplines. The technical details of the communication specification can be found in the ADPC specification (see \cite{humanAdvancedDataProtection2021}) and have been reported in other academic publications.

\paragraph{Approach:} The ADPC was developed through a collaboration of an interdisciplinary team of computer scientists, social scientists, lawyers, activists, and practitioners, as a part of the RESPECTeD project\footnote{\url{https://respected.eu}} led by the Soheil Human (Sustainable Computing Lab) and Max Schrems (NOYB – European Center for Digital Rights). In this paper, we present the ADPC based on the questions that were identified through a set of interdisciplinary qualitative studies (a mixture of two 
focus groups and eight 
interviews) in which sixteen
experts working on data protection and consenting topics with diverse expertise and backgrounds (from computer science, natural science, law, social science and humanities) participated. The studies included other aspects that are not reported in this paper. %
We hope that the provided \textit{questions-based} narrative can make it easier for a non-technical audience to understand the bases of the ADPC.

\paragraph{Structure:}
In the following, we first reflect on some of the most critical problem areas that the ADPC aims to tackle--which will also provide a brief overview of the state of the art. Then we will present the ADPC to an interdisciplinary audience through the questions identified in our qualitative studies. Finally, we discuss the next steps and conclude this paper by reminding the importance of \textit{Data Protection and Consenting Communication Mechanism} (DPCCMs) \cite{Human2022DPCCM} such as the ADPC.

%% file: 02_problems-and-necessities.tex
\section{The Necessity of Advanced Mechanisms for Data Protection and Consenting Control Mechanisms}
\label{sec:necessity}

In our increasingly digital world, data in general, and personal data in particular, has become an important driver of digital innovation and economic growth \cite{jetzek2014data, manyika2011big, sorescu2017data}, causing \textit{digitization} and \textit{digital transformation} in many sectors and application areas \cite{2022__HumanEtAlCallInterdisciplinaryResearch}.
Keeping this \textit{transformation} sustainable, however, has been a challenging mission \cite{2022__HumanEtAlCallInterdisciplinaryResearch}.
Among others, the tension between providing personalized services --- used in many cases as a basis for personalized advertisements --- and respecting humans' privacy or agency has remained unresolved.  

In the following, we argue that advanced Data Protection and Consenting Communication Mechanisms (like the ADPC) can contribute to resolving some of the most fundamental existing challenges concerning online privacy and consenting. Clearly, many different systemic, structural, technical, societal, economic, or political reasons are involved in the co-construction of the current and the future \textit{digital world}.%
Therefore, the ADPC can only be helpful if it is well supported by other sociotechnical means (e.g., complementary technical solutions, regulations, standards, policies, business models, or ethical frameworks).

\subsection{Who is Controlling the \textit{Data} Concerning Privacy and Consenting?}
According to different legal frameworks such as the European General Data Protection Control (GDPR), data controllers are obliged to provide data subjects with privacy and consenting related information and --- whenever needed --- to obtain their consent or privacy-related decisions through lawful and ethical means. This is currently mainly practiced through the mechanisms that the data controllers provide (e.g., so-called ``cookie banners'') \cite{degeling2018we}. Even if designed in transparent, understandable, ethical, and lawful manners---which is often not the case (see e.g., \cite{utz2019informed, humanHumanCentricPerspectiveDigital2021, santos2019cookie})---since the data-controllers design, develop, maintain, and provide these mechanisms (interfaces) on their sides (e.g., on their websites or apps), the data related to users' privacy and consenting, e.g., their consent or decisions, \textit{remains solely under the control of the data controllers}. This means that the data subjects do not even receive a copy of the data or any confirmation (e.g., receipts, see also \cite{jesus2020towards}) related to their decisions (or related to the information that was provided to them). This puts the data subjects (i.e., humans, users, customers, etc.) in an inherently weak position. As described in the literature (e.g., \cite{humanHumancentricPersonalData2022}), humans have difficulties managing their online data protection and consenting from different cognitive, collective, and contextual perspectives. Without having \textit{access to data}, empowering \cite{2020__HumanEtAlEndUserEmpowerment} humans with the expected cognitive, collective, or contextual supports \cite{humanHumancentricPersonalData2022} \textit{on their own side (on the user side)} is almost impossible. 
The ADPC makes it possible to communicate such data in a bidirectional manner between different involved actors, solving one of the most essential issues of the current practice of personal data processing on the Internet.

\subsection{Who is Controlling the \textit{Procedures} concerning Data Protection and Consenting?}
Along with controlling privacy and consenting data, data controllers currently fully control the data protection and consenting \textit{procedures}. For example, the data controllers decide when the \textit{consent pop-ups} appear or what kind of decisions should be communicated. This makes it possible for them to design the procedures in favor of their own interests. For example, consenting to all is the ``default easy option'' that can be performed through a \textit{consent banner}, and withdrawing a previously given consent needs a time-consuming procedure using different forms that are sometimes hidden and hard to find. The ADPC makes it possible for the data subjects to have autonomy and agency concerning the \textit{data protection and consenting procedures} and, if needed, start the procedure themselves or modify the offered procedures. For example, a data subject can start the communication procedure themself by sending a withdrawal message to a data controller, or if a data controller sends different requests, the data subject can only answer to a subset of them. This provides a \textit{new balance} between the data subjects and data controllers. %

\subsection{Who is Controlling the \textit{User Interfaces and Designs} Concerning Data Protection and Consenting?}
In line with the previous two points, data controllers can currently decide on every detail of the user interfaces used in privacy and consent-obtaining mechanisms. This has increased the use of negative nudging mechanisms (also sometimes called \textit{dark patterns}) in the \textit{consent banners} \cite{nouwens2020dark}. The ADPC brings privacy and consenting user interfaces to the user side, putting them under the control of the users and/or their trusted supporting actors (e.g., trusted family, friends, experts, NGOs, privacy-preserving technology developers, or browsers). In other words, complemented with other sociotechnical solutions, the ADPC can contribute towards ending (or at least reducing) the issue of \textit{dark patterns} in relation to online privacy and consenting on the Internet --- among others.

\subsection{Empowering Data Subjects, Empowering Data Controllers}
The ADPC can empower users by involving them in controlling procedures, designs, and data management of online personal data processing. But, besides data subjects --- that have difficulties dealing with online privacy and need to be empowered ---, complying with different legal frameworks and developing various privacy management mechanisms is not an easy task for many data controllers as well (see, e.g., \cite{mikkelsen2019gdpr, tsaneva2019challenges}). Among others, developing consent mechanisms (e.g., so-called \textit{cookie banners}) can be challenging, especially for smaller companies. The ADPC can enable novel means of communicating privacy and consenting with data subjects that might reduce the load of designing and maintaining some of the existing mechanisms, such as the \textit{cookie banners}. Moreover, the current practice can be very disturbing for users. Considering that \textit{user experience} is a significant factor for online service providers, replacing the banners with more advanced mechanisms that the ADPC can enable can be very beneficial for the companies \cite{salutari2020implications}.

%% file: 03_adpc.tex
\section{The Advanced Data Protection Control}
\label{sec:ADPC}
In the previous section, we argued that the lack of widely used, \textit{bidirectional} and \textit{human-centric} communication mechanisms concerning privacy and consenting data, meta-data, information, requests, preferences, and decisions --- together with other reasons such as the data controllers' or browser companies' business models or conflicts of interest --- has led to a mainstream practice of online privacy and consenting in which most of the fundamental aspects such as \textit{data}, \textit{procedures}, and \textit{user-interfaces} are designed, developed, maintained and controlled by the data controllers. This has caused an unfair and uneven power balance which can have different societal consequences \cite{humanHowCanPluralist2019}. The ADPC aims to tackle this fundamental issue by realizing the lacking communication mechanism.
However, as a sociotechnical solution that should be complemented by other interdisciplinary means, the bases of the ADPC need to be communicated in an easy-to-understand manner for wider non-technical experts to enable the development of future multidisciplinary solutions through the collaboration of experts from different fields. To this end, as mentioned before, we used a set of qualitative studies to identify some of the main interdisciplinary questions concerning the ADPC and aligned this introductory paper based on these questions.  In the following, the main identified questions are listed and briefly answered: 

\subsection{What is the ADPC?}

The Advanced Data Protection Control (or the ADPC) is a technical specification (and the sociotechnical solutions surrounding it) for the communication of data protection and consenting data, meta-data, information, requests, preferences, and decisions. For example, the ADPC specification defines automated means for data subjects (e.g., website visitors) to 1) give or refuse consent for the specific purposes that the data controller describes, 2) withdraw any consent already given, as well as 3) object to processing for direct marketing purposes based on the data controller’s legitimate interest. This enables the user to easily manage data protection decisions through the web browser and possibly customize how requests are presented and responded to (e.g., using a browser extension to import or specify lists of trusted websites). The result could be comparable to how websites ask for permission to access a webcam or microphone via a browser: the browser keeps track of the user’s decisions on a site-by-site basis, ensures that the user gets a genuinely free choice (e.g., no \textit{dark pattern} used), and puts the user in control over their decisions.

\subsection{How does ADPC work?}

APDC specification defines a mechanism for expressing user decisions about personal data processing under the European Union's data protection regulations and similar regulations outside the EU. The mechanism functions by exchanging HTTP headers between the user agent and the web server or through an equivalent JavaScript interface. In the future, other protocols and technologies can also be used to transfer \textit{ADPC signals}.

The mechanism serves as a means for users to give or refuse consent, withdraw any consent already given, and object to processing --- among others. The mechanism provides an alternative to existing non-automated consent management approaches (e.g., \textit{cookie banners}) and aims to reduce the efforts of the different parties involved regarding the protection and management of users' privacy.

\subsection{What are the legal foundations of the ADPC?}
Besides the Charter of Fundamental Rights of the European Union, as it is mentioned in the ADPC specification \cite{humanAdvancedDataProtection2021}, different legal frameworks such as the European Union's General Data Protection Regulation (GDPR) and ePrivacy Directive define rights and obligations around the processing of personal data\footnote{The ADPC can also be used to manage privacy and consenting based on non-European legal frameworks}. The starting position of the GDPR is that the processing of personal data is only lawful if it has an appropriate legal basis; one basis being that ``the data subject has given consent to the processing of his or her personal data for one or more specific purposes'' (point (a) of Article 6(1) GDPR). Similarly, the ePrivacy Directive (in Article 5(3)) requires the user's consent when any data is stored on or retrieved from terminal equipment beyond what is strictly necessary. Moreover, when a data controller relies on legitimate interest as the legal basis for direct marketing, the user has an absolute right to object under Article 21(2) GDPR.

As website publishers often desire to process their visitors' personal data for purposes beyond what is necessary to serve the website and beyond what can be based on legitimate interest, a website operator often wants to ask whether the visitor consents to such processing. Such communication currently tends to be done via highly disruptive and repetitive interfaces contained in the web page itself (e.g., \textit{cookie banners}) rather than through the web browser or other automated channels.

It is the user's choice how to communicate the exercise of GDPR rights to a data controller --- the user could send an email, letter, or click a button on a website. In addition, technical means can be used:

Article 21(5) GDPR expressly provides that ``the data subject may exercise his or her right to object by automated means using technical specifications``.
Recital 32 of the GDPR also makes clear that requesting and giving consent could take many forms, which ``could include ticking a box when visiting an internet website, [or] choosing technical settings for information society services […]'', as long as it satisfies the requirements such as being informed and unambiguous.
Recital 66 of Directive 2009/136/EC, which updates the 2002 ePrivacy Directive, likewise states that ``the user's consent to processing may be expressed by using the appropriate settings of a browser or other application''.
The proposed ePrivacy Regulation (2017/0003 (COD)) equally foresees automated means to communicate data subject preferences.
Despite various legal provisions suggesting its validity, a standardized means for communicating GDPR rights has thus far been lacking.

\subsection{What does make the ADPC ``\textit{Advanced}''?}

There have been other attempts to implement automatic privacy controls (such as the ``Do Not Track''\footnote{https://www.w3.org/TR/tracking-dnt/} (DNT) and its recently adapted revision the ``Global Privacy Control''\footnote{https://globalprivacycontrol.github.io/gpc-spec/} (GPC) \cite{zimmeck2020standardizing}). The ADPC is different because it has been designed to better integrate with the requirements of GDPR and the upcoming ePrivacy Regulation, as well as with other international laws:
\begin{itemize}

\item The ADPC is domain-specific (`site specific'), so users can choose to tailor their interaction with different websites and data controllers.
\item The ADPC allows opt-in (consent) and opt-out (objection) signals, whereas other signals were based on an opt-out framework.
\item  The ADPC allows domains to freely define a consent request or use a formulation standardized by industry groups (like the IAB's TCF specification). This makes ADPC open and interoperable with other systems.
\item The ADPC allows general signals (like ``reject all'', ``withdraw all'', ``object to all'', ``do not track'', ``do not sell''), specific signals (like consent to a specific request) and a combination of general and specific signals (like ``reject all, but consent to requests `x' and `y''').
\item  The ADPC allows browsers, plugins, or operating systems to provide users with settings and logic that determines how requests are treated. This includes white- and blacklisting, industry-wide purposes, or logic like showing a request only when visiting a page regularly.
\item  The ADPC limits the (legal) fingerprinting surface by not sending any signal if a domain does not support The ADPC (and thereby publicly commits not to use the signal further), as well as sending different signals to different domains.

\end{itemize}

\subsection{Does the ADPC provide its own vocabulary?}
The ADPC is not limited to any vocabulary (ontology). It can be used with different vocabularies depending on the sector, use case, legal requirements, etc. However, the ADPC can be complemented with standardized vocabularies, e.g., \cite{pandit2019creating}. 

\subsection{Who can start the procedure?}
As discussed in the previous section, having the \textit{power} of starting the procedure of communicating privacy and consenting data and decisions is an essential factor that shapes the dynamics of online personal data processing. \textit{Privacy signals} such as the DNT or the GPC make it possible for the users to send a single binary message to data controllers regardless of data controllers' consent obtaining mechanisms. Contrary to them, the current consent obtaining mechanisms, such as the \textit{cookie banners}, give full control of starting the procedures to data controllers. The ADPC, however, provides a bidirectional mechanism that allows each party
to start the communication. For example, a data subject can send a withdrawal request without waiting for the data controllers to send a query, or a data controller can send a set of requests to a data subject to start the procedure ––– both are possible. %

\subsection{Who decides about the user interface design?}
\begin{figure*}
    \centering
  \includegraphics[width=0.7\textwidth]{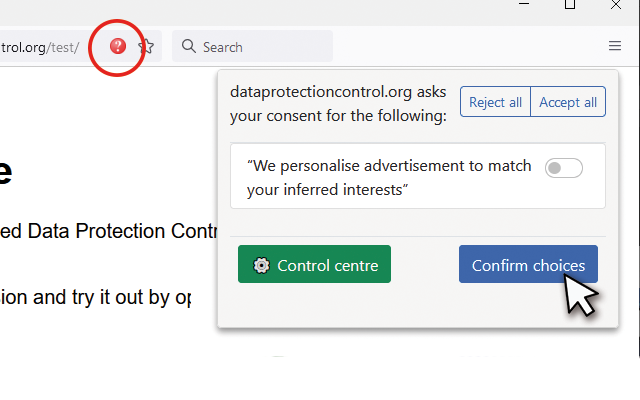}
  \caption{An example of a user-side ADPC-based GUI that is shown in a browser under the control of a data subject}
  \label{fig:ADPC}
    \label{fig:adpc-ui}
\end{figure*}

A big advantage of the ADPC is that it brings the representation and decision-making mechanisms to the \textit{user-side}. Depending on the implementation, the user themself, the browser companies, the plugin developers, or trusted actors can decide about (or design) the representation and decision-making mechanisms. As a result, the ADPC-based user interfaces, if designed in a Human-centric, Accountable, Lawful, and Ethical (HALE \cite{humanHALEWHALEFramework2022}) manner, can reduce (or eliminate) the usage of problematic nudging mechanisms (e.g., so-called \textit{dark patterns}) in privacy-related solutions, by shifting the control from the data controllers to data subjects (or their trusted parties). Figure~\ref{fig:adpc-ui} shows an example of an ADPC-based graphical user interface (GUI) embedded in a browser under the control of a data subject. 

\subsection{What about supporting users?}
While bringing the \textit{privacy and consenting data} to the \textit{user-side} and empowering users through involving them in controlling the procedures and designs of personal data processing is certainly very important, as it is discussed in the literature \cite{humanHumanCentricPerspectiveDigital2021,humanHowCanPluralist2019,humanHumancentricPersonalData2022}, users--- \textit{as human-beings}---have limited cognitive capacities, knowledge, expertise, time, or motivation to manage their privacy \textit{alone}. They need to be \textit{empowered} \cite{2020__HumanEtAlEndUserEmpowerment} by sociotechnical means, such as the Personal Data Protection and Consenting Assistant Systems (PDPCAS) \cite{humanHumancentricPersonalData2022}, that provide \textit{cognitive}, \textit{collective}, and \textit{contextual} (\cite{kirchner2019context, humanContext2021}) supports for them. Such systems can provide users with, e.g., automation tools, management tools, memorizing tools, trust assessment tools. Moreover, by being empowered to control the procedures, users can be supported through \textit{white lists} or \textit{black lists} that can help them (or their \textit{agents, i.e., their PDPCASs}) to interact with online services or make decisions regarding their privacy in easier manners. Without the ADPC (or similar mechanisms) development of PDPCASs or other supporting tools is almost impossible since such systems need to have access to privacy and consenting data and be involved in the \textit{procedures} to function.

\subsection{Is the ADPC only limited to the web?}
Currently, the ADPC supports HTTP and JavaScript, which means that it can be used in browsers and potentially web apps, mobile apps, and other solutions (from smart TVs to different IoT devices) that are based on these technologies. However, the ADPC aims to support other technologies as well in the future. For example, the Sustainable Computing Lab is currently leading a project that aims to bring the ADPC to IoT devices using \textit{Bluetooth}.

\subsection{What is the difference between the ADPC and GPC or DNT?}
Do Not Track (DNT) and Global Privacy Control (GPC) are binary HTTP header \textit{signals} that are developed based on an approach to online privacy that is closer to California's legal framework, such as the California Consumer Privacy Act (CCPA). The ADPC, on the other hand, is a \textit{bidirectional advanced communication mechanism} that can be used to communicate many different types of information and decisions related to privacy and consenting. The ADPC can be implemented in a way to produce binary signals similar to DNT or GPC, but it is not limited to that. While the ADPC was designed with European laws in mind,  it is not limited to any legal frameworks and can be adapted to them.

\subsection{What is the benefit for the data controllers?}
As it was mentioned before, several benefits can be considered for the data controllers. Among others: 1) supporting the ADPC can show that companies respect their users' privacy and agency. This might be seen as a value-proposition (see e.g., \cite{simkevitz2009privacy}) for many of the companies and increase their trustworthiness.
2) Developing \textit{consent banners} and maintaining them can be a costly and difficult task. The ADPC can reduce the load on data controllers in this regard.
3) The current \textit{consent banners} are very distracting and reduce visitors' or customers' \textit{user experience}. The ADPC can support the companies to eliminate them and improve their users' satisfaction and revenue (see, e.g., \cite{salutari2020implications}).

\subsection{What about if someone misuses the ADPC?}
Security and privacy measures have been considered in the development of the ADPC. For example, the ADPC is domain-specific (see above), which can highly reduce potentially problematic fingerprinting. But similar to other technologies, the ADPC might also be misused. Depending on the sophistication of the specific implementations, the developers are expected to consider further privacy and security measures when implementing ADPC-based technologies. An analogy that can be helpful here is the \textit{email protocols} (e.g., the Simple Mail Transfer Protocol, SMTP): while SMTP was designed by considering specific privacy and security measures, it does not prevent people from sending \textit{spams}:  other complementary solutions that provide different \textit{anti-spam techniques
} are expected to be included in the systems that implement SMTP. This is also the case with the ADPC; privacy and security measures and complementary solutions should be implemented along with the ADPC, depending on the application area, use case, and underlying technological systems.

%% file: 06_conclusion.tex
\section{Conclusion}
\label{sec:conclusions}
In this paper, based on an expert study, we identified some of the essential questions that need to be addressed regarding the Advanced Data Protection Control (ADPC). We used these questions as a basis to introduce ADPC to interdisciplinary audiences that might not have deep technical expertise. We believe that the APDC has the potential to shift the power structure of the Internet, and it is essential to communicate its different aspects with a wide range of scientists, practitioners, activists, businesses, and policy-makers. We are also well aware that no technology alone can be impactful if other socio-technical solutions will not support it. We call interdisciplinary communities that are working on privacy and consenting to work together to co-construct a \textit{next generation Internet} that is Human-centric, Accountable, Lawful, and Ethical and realizes digital Sustainability as an essential aspect of our societies. 

%% file: 07_acknowledgement.tex
\section*{Acknowledgement}
On behalf of the ADPC core team (Soheil Human, Max Schrems, Alan Toner, and Gerben), I acknowledge all people who contributed to the development of the ADPC. Many different colleagues and friends from the Vienna University of Economics and Business, Sustainable Computing Lab, NOYB, W3C community groups, various data protection boards and authorities, universities and research institutions, NGOs, parliaments and governments, international organizations, standardization bodies, journals and news agencies, open source developers and communities, and companies supported this project by providing us with their input and feedback. We are very thankful for all these contributions. The ADPC was partially funded by \textit{netidee} program of the ``Internet Privatstiftung Austria – Internet Foundation Austria'' (grant number \textit{prj4625}).